\documentclass{pasj00}

\begin{document}

\SetRunningHead{K. Niinuma et al.}{Astrometry of IRAS 06061+2151}
\Received{2010/07/23}
\Accepted{2010/09/21}

\title{Astrometry of H$_2$O masers in the massive star--forming region IRAS 06061+2151 with VERA}

\author{
Kotaro \textsc{NIINUMA},\altaffilmark{1}
Takumi \textsc{NAGAYAMA},\altaffilmark{1}
Tomoya \textsc{HIROTA},\altaffilmark{1, 2}
Mareki \textsc{HONMA},\altaffilmark{1, 2}
Kazuhito \textsc{MOTOGI},\altaffilmark{6}
Akiharu \textsc{NAKAGAWA},\altaffilmark{5}
Tomoharu \textsc{KURAYAMA},\altaffilmark{5}
Yukitoshi \textsc{KAN-YA},\altaffilmark{1}
Noriyuki \textsc{KAWAGUCHI},\altaffilmark{1, 2, 3}
Hideyuki \textsc{KOBAYASHI},\altaffilmark{1, 4}
Yuji \textsc{UENO},\altaffilmark{2}
}

\altaffiltext{1}{Mizusawa VLBI Observatory, National Astronomical Observatory of Japan, 2-21-1 Osawa, Mitaka, Tokyo 181-8588}
\altaffiltext{2}{Graduate University for Advanced Studies, 2-21-1 Osawa, Mitaka, Tokyo 181-8588}
\altaffiltext{3}{Mizusawa VLBI Observatory, National Astronomical Observatory of Japan, 2-12 Hoshi-ga-oka, Mizusawa-ku, Oshu, Iwate 023-0861}
\altaffiltext{4}{Department of Astronomy, The University of Tokyo, 7-3-1 Hongo, Bunkyo-ku, Tokyo 113-8654}
\altaffiltext{5}{Graduate School of Science and Engineering, Kagoshima University, 1-21-35 Korimoto, Kagoshima, Kagoshima 890-0065}
\altaffiltext{6}{Department of Cosmosciences, Graduate School of Science, Hokkaido University, N10 W8, Sapporo 060-0810}

\email{kotaro.niinuma@nao.ac.jp}


%


\maketitle

\begin{abstract}
We measured the trigonometric annual parallax of H$_2$O maser source associated with the massive star-forming regions IRAS 06061+2151 with VERA. The annual parallax of $0.496\pm0.031$ mas corresponding to a distance of $2.02^{+0.13}_{-0.12}$ kpc was obtained by 10 epochs of observations from 2007 October to 2009 September. This distance was obtained with a higher accuracy than the photometric distance previously measured, and places IRAS 06061+2151 in the Perseus spiral arm. We found that IRAS 06061+2151 also has a peculiar motion of larger than 15 km s$^{-1}$ counter to the Galactic rotation. That is similar to five sources in the Perseus spiral arm, whose parallaxes and proper motions have already been measured with higher accuracy. Moreover, these sources move at on average 27 km s$^{-1}$ toward the Galactic center and counter to the Galactic rotation.
\end{abstract}


\section{Introduction}
The annual parallax is the most direct way to accurately measure the distance to a target source without any assumption. Very Long Baseline Interferometer (VLBI) provides the highest accuracy of position determination among astronomical instruments. It is possible to determine the position of target sources with an accuracy of less than 1 mas with the VLBI observations. The phase-referencing VLBI has been utilized in recent astrometry to measure the annual parallaxes of target sources. For example, the results of trigonometric parallax measurements of H$_2$O and CH$_3$OH maser sources using the NRAO Very Long Baseline Array (VLBA) were reported by Xu et al. (2006); Hachisuka et al. (2006, 2009); Reid et al. (2009a).

The Japanese VLBI Exploration of Radio Astrometry (VERA) is an instrument dedicated to carrying out VLBI astrometry. It aims to figure out the Galactic 3D structure by measuring the trigonometric parallaxes and the proper motions of maser sources with the phase-referencing VLBI. To carry out effective phase-referencing observations, the dual-beam system was developed \citep{kobayashi08}. The results of the trigonometric parallaxes of several H$_2$O maser source (e.g. Orion-KL, IRAS 06058+2138) with VERA have already been reported by Choi et al. (2008); Hirota et al. (2007, 2008a, 2008b); Honma et al. (2007); Imai et al. (2007); Nakagawa et al. (2008); Oh et al. (2010); Sato et al. (2008, 2010).

It is believed that our Galaxy is composed of several spiral arms. In order to understand the spiral structure and the rotational kinematics of the Galaxy, it is important to study the structure of each spiral arm. So, measuring the accurate distance to the astronomical objects and investigating their kinematics in the spiral arms are essential for studying spiral arm structure. The Perseus spiral arm is the nearest one to us outward from the Sun in the Galaxy and consists of many star-forming regions. Also, the distances to six sources have previously been measured with high accuracy in previous studies \citep{moellenbrock09, moscadelli09, oh10, reid09a, sato08, xu06}. It turns out that some of these sources (e.g. W3OH, IRAS 00420+5530) showed their peculiar motion with respect to the Galactic rotation. However, it is necessary to increase the number of samples because the current number is insufficient to indicate that their peculiar motions relate to the motion of the Perseus spiral arm.

\begin{table*}[t]
\caption{VERA observations of IRAS 06061+2151}
\label{tab:tbl1}
\centering
\begin{tabular}{clcccccc} \cline{5-8}
\hline
Epoch ID & \multicolumn{1}{c}{Date} & Day of year & Synthesized beam & \multicolumn{4}{c}{$T_{\mathrm{sys}}$} \\\cline{5-8}
 & \multicolumn{1}{c}{} &  & $\theta_{maj}\times\theta_{min} (PA)$ & Iriki & Mizusawa & Ogasawara & Ishigakijima \\
 &                                 &  &    [mas$\times$mas (\timeform{D})] & \multicolumn{4}{c}{[K]}\\\hline\hline
1 & 2007/Oct 23 & 2007/296 & $1.80\times1.12$ (128.5) & 150 - 240 & 110 - 150 & 320 - 600 & 240 - 550 \\
2 & 2007/Dec 15 & 2007/349 & $1.83\times1.13$ (127.4) & 110 - 170 & 110 - 530 & 150 - 210 & 250 - 550 \\
3 & 2008/Feb 15 & 2008/046 & $1.81\times1.13$ (128.9) & 100 - 160 & 100 - 190 & 150 - 2600 & 150 - 300 \\
4 & 2008/Apr 20 & 2008/111 & $1.84\times1.13$ (127.4) & 130 - 200 & 130 - 200 & 160 - 270 & 250 - 550 \\
5 & 2008/Jun 15 & 2008/167 & $1.79\times1.12$ (128.5) & 500 - 12000 & 180 - 260 & 300 - 550 & 120 - 200 \\
6$^*$ & 2008/Sep 29 & 2008/273 & $1.74\times1.09$ (129.7) & - & 120 - 200 & 450 - 850 & 350 - 1500 \\
7 & 2008/Nov 02 & 2008/307 & $1.86\times1.13$ (125.7) & 250 - 2300 & 140 - 280 & 200 - 380 & 300 - 750 \\
8 & 2008/Dec 14 & 2008/349 & $1.90\times1.13$ (133.4) & 100 - 140 & 100 - 130 & 200 - 380 & 180 - 1400 \\
9 & 2009/Feb 15 & 2009/046 & $1.86\times1.14$ (130.7) & 150 - 700 & 110 - 180 & 240 - 5000 & 210- 540 \\
10 & 2009/May 11 & 2009/131 & $1.82\times1.13$ (128.5) & 100 - 145 & 200 - 340 & 150 - 240 & 200 - 450 \\
11 & 2009/Sep 07 & 2009/250 & $1.89\times1.15$ (128.5) & 190 - 360 & 250 - 580 & 290 - 600 & 210 - 440 \\\hline
\multicolumn{8}{@{}l@{}}{\hbox to 0pt{\parbox{150mm}{\footnotesize 
*: We do not include the data of the epoch 2008/273 for the astrometric analysis (Sect. \ref{observation}).

**:In the column of $T_{\mathrm{sys}}$, we list the minimum and the maximum of $T_{\mathrm{sys}}$ during the observation at each station and in each epoch.

}\hss}}

\end{tabular}

\end{table*}

IRAS 06061+2151 is a massive star-forming region located in the Gemini OB1 cloud complex. Several sources are associated with IRAS 06061+2151, for example, a radio continuum source associated with the ultra compact H\emissiontype{II} region, G188.794+1.031 \citep{kurtz94}, a sub-millimeter wavelength source JCMTSF J060907.1+215037 \citep{difrancesco08}, a near infrared (NIR) cluster GL 5182, the C$^{18}$O line emission \citep{saito07}. Moreover, this region shows the H$_2$O maser emission \citep{kompe89}. Thus the structure and the kinematics of a molecular gas around this region have been well studied with high spatial resolution observations. Especially, \citet{motogi08} found the bipolar morphology and the expanding motion suggested by the relative proper motion measurement of H$_2$O masers with the Japanese VLBI Network (JVN: the maximum angular resolution is about 1 mas at 22 GHz). The photometric distance to the Gemini OB1 cloud complex was measured by  the previous studies, and the distance range was from 1.5 kpc to 3.5 kpc \citep{carpenter95,humphreys78,moffat79}. IRAS 06058+2138 whose parallax distance has already been measured as approximately 2 kpc by VLBI observation, is known to be located in the Perseus spiral arm and located near IRAS 06061+2151 on the celestial  sphere. Therefore, although the uncertainty of the photometric distance is much larger, there is a possibility that IRAS 06061+2151 may also be located in the Perseus spiral arm.

In this paper, we report on the result of the annual parallax measurement of the H$_2$O maser source in IRAS 06061+2151 with VERA, and discuss the motion in the Galaxy.

\begin{figure*}[t]
  \centering
    \FigureFile(140mm,){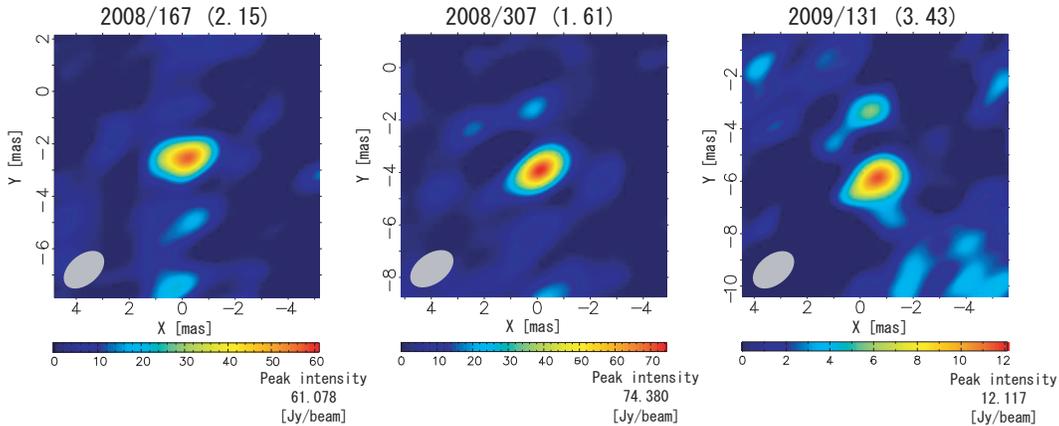}
  \caption{The phase-referenced map of the maser spot at $v_{\mbox{\scriptsize LSR}}=-7.795$ km s$^{-1}$ in three epochs. Epochs when the maser spot was observed and the ratio of the maser spot size whose flux has more than 30\% of brightest component in the map to beam size are shown on the top of each map, as 'DOY (spot size)'. The most right figure was not included in our result because spot size of the maser spot exceeds the threshold for conducting an astrometric analysis in the presented study (see Sect. \ref{parallax}, and \ref{spotsize}).}\label{fig:f1}
\end{figure*}

\section{VLBI observations with VERA}\label{observation}
VERA observations of the H$_2$O maser source ($6_{16}-5_{23}$, 22235.080 MHz) associated with IRAS 06061+2151 have been carried out since 2007 October to 2009 September at 2- to 3-month intervals. We present the results from a total of 11 observing epochs: 2007/296, 2007/349, 2008/046, 2008/111, 2008/167, 2008/273, 2008/307, 2008/349, 2009/046, 2009/131, and 2009/250; hereafter an observing epoch is denoted by year/day of the year (DOY). All 4 VERA stations, providing a maximum baseline length of 2270 km, participated in 10 epochs, but only 3 VERA stations (Mizusawa, Ogasawara, and Ishigakijima: see Figure \ref{fig:f1} of \citet{petrov07}) participated in the 2008/273 epoch. Therefore, we do not include the data of the epoch 2008/273 in order to conduct a more accurate astrometric analysis.

Observations were made in the dual-beam mode and we observed simultaneously \citep{kawaguchi00, honma08a} the H$_2$O maser source in IRAS 06061+2151 (J2000.0 $\alpha=\timeform{06h09m06s.97458}\pm\timeform{0s.006}$, $\delta=\timeform{+21D50'41".4045}\pm\timeform{0".05}$; Migenes et al. 1999), and the reference source J0603+2159 (J2000.0 $\alpha=$ \timeform{06h03m51s.55709}$\pm$\timeform{0s.00008}, $\delta=$ \timeform{+21D59'37".6976}$\pm$\timeform{0".0035}; Fey et al. 2004). The separation angle between those two was \timeform{1D.22}. J0603+2159 was detected with a peak flux density ranging from about 60 to 140 mJy beam$^{-1}$ throughout all epochs. Left-handed circular polarization was received and sampled with 2-bit quantization and filtered using the VERA digital filter unit \citep{iguchi05}. The data were recorded onto magnetic tapes at a rate of 1024 Mbps, providing a total bandwidth of 256 MHz which was divided into 16 IF channels (16 MHz each). 8 MHz of the one IF was assigned to IRAS 06061+2151 and the other 15 IFs were assigned to J0603+2159. A bright continuum source, DA193, was observed every 90 minutes for bandpass and delay calibration. The system noise temperatures ($T_{\mathrm{sys}}$) including atmospheric attenuation were depending on the weather conditions and the elevation angle of the observed sources. We list the value of $T_\mathrm{sys}$\footnote{We list the minimum and the maximum of $T_{\mathrm{sys}}$ during the observation at the each station and the each epoch in Table \ref{tab:tbl1}.} as well as the typical synthesized beam sizes (major and minor axes) and their position angles at all epochs in Table \ref{tab:tbl1}. The aperture efficiencies of the antennas ranged from 45 to 52 \%, depending on the stations (see, the VERA status report\footnote{http://veraserver.mtk.nao.ac.jp/restricted/CFP2009/status09.pdf}). A variation of the aperture efficiency of each antenna as a function of the elevation angle was confirmed to be less than 10 \%, even at the lowest elevation in the observations ($\sim$\timeform{20D}). Correlation processing was carried out on the Mitaka FX correlator \citep{chikada91} on the NAOJ Mitaka campus. For H$_2$O maser lines, the spectral resolution was set at 15.625 kHz, corresponding to a velocity resolution of 0.21 km s$^{-1}$. So, the bandwidth of 8 MHz corresponds to the velocity coverage of 107.5 km s$^{-1}$.

\section{Data reduction with VEDA}\label{reduction}
We conducted the data reduction with the software package VERA Data Analyzer (VEDA). VEDA has been developed for the astrometric analysis of data observed with the VERA dual-beam system by the software development group in Mizusawa VLBI observatory of NAOJ.

At first we conducted the averaging of time and frequency for the data of the position reference source J0603+2159 in order to reduce the data size. After data averaging, we carried out the amplitude calibration and the bandpass calibration for the maser source in IRAS 06061+2151 and J0603+2159 independently. The amplitude calibrations were performed based on the ($T_{\mathrm{sys}}$) data monitored during the scan. The bandpass calibrations were performed utilizing auto-correlation spectra of the blank sky data and the bright calibrator source (DA193).

The fringe search was conducted with the visibility data of the fringe finder DA193 in order to determine the delay and the delay-rate offset. Next the fringes were searched for J0603+2159 by utilizing the station solution determined by DA193, and the image of J0603+2159 was obtained by the deconvolution (CLEAN) and the self-calibration algorithm. 

The absolute position of the maser source was obtained by applying the phase solutions of J0603+2159 to the maser source (phase-referencing method). For the maser source, the Doppler tracking were made. Finally, we made images for each velocity channels, which were selected so that LSR velocities of maser features reported in \citet{motogi08} are fully covered, in order to search for maser spots. Each of the velocity channels was imaged with a field of view of $2\timeform{"}\times 2\timeform{"}$ with 1024 grids $\times$ 1024 grids by CLEAN with a threshold for signal to noise ratio of 7 ($7\sigma$ detection limit). For each detected maser spot, the images of 10 mas $\times$ 10 mas were obtained with 512 grids $\times$ 512 grids. 

After the imaging process, we conducted the astrometric analysis using the result of phase-referenced analysis. For the astrometric analysis, the maser spots without extended structures should be selected so as to accurately determine the position. So, we can set the threshold for the structure of maser spot using the parameter "spot size" with VEDA. The spot size is defined as the ratio of an area whose flux is larger than 30 \% of the peak flux density to the synthesized beam size. We set the limit of 5.0 for the spot size in imaging process, so the maser spots whose size extend beyond 5.0 times the beam size were rejected. As an example, we show channel maps of the same spot detected with a different spot size at several epochs in Figure \ref{fig:f1}. We discussed the details of the spot size in Sect. \ref{spotsize}.

\section{Results}\label{result}
\begin{figure}[h]
  \centering
    \FigureFile(75mm,){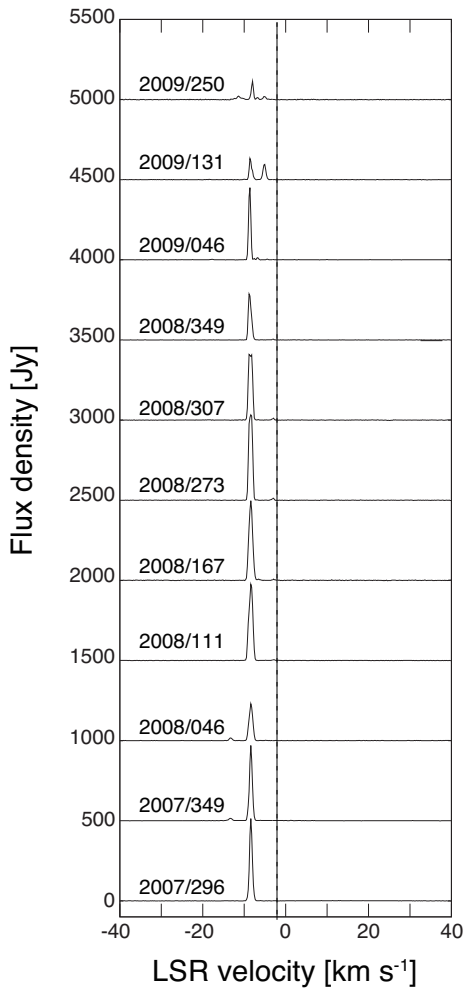}
  \caption{Scalar averaged cross power spectra of H$_2$O maser emission toward IRAS 06061+2151 obtained using VERA. The spectra are integrated for 900 s at the Mizusawa-Ishigakijima baseline. The dashed line indicates the systemic velocity of -1.6 km s$^{-1}$ \citep{wouterloot89}.}\label{fig:f2}
\end{figure}

\begin{figure*}[t]
  \centering
    \FigureFile(140mm,){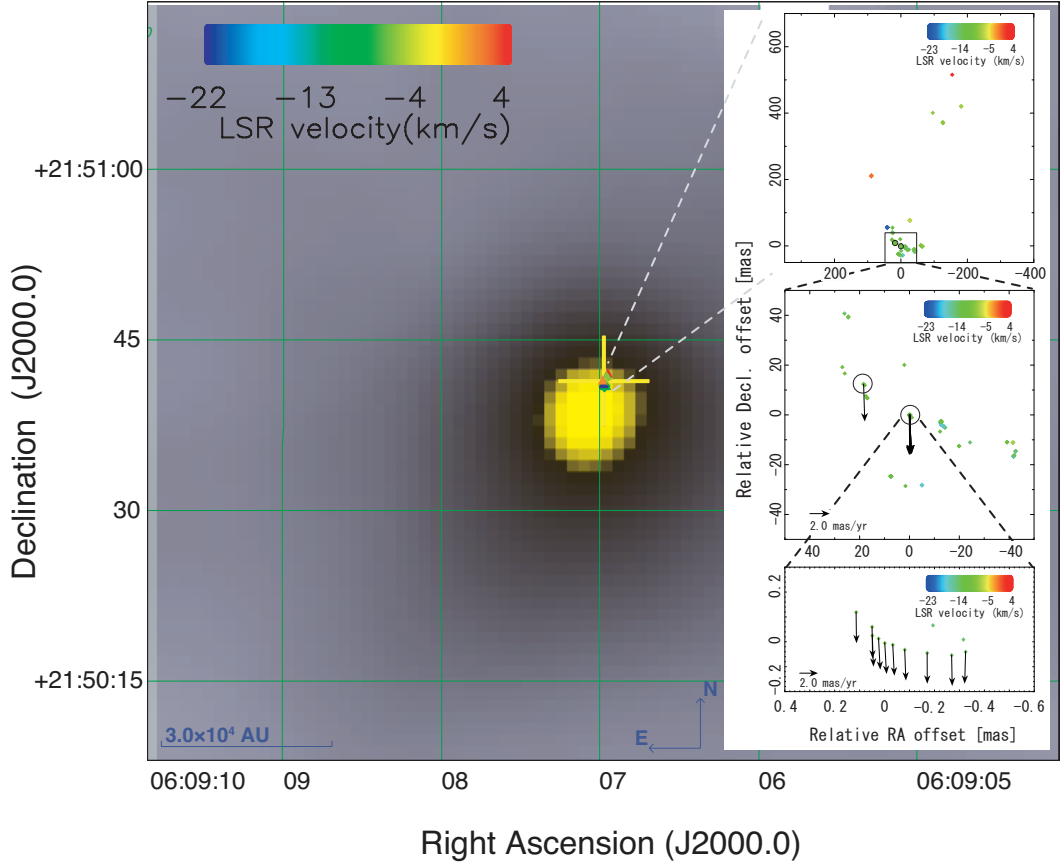}
  \caption{850 $\mu$m SCUBA image (gray scale), the yellow cross indicates the center position of 8.4 GHz continuum image observed with VLA B-configuration \citep{kurtz94}, and the distribution of the H$_2$O maser spots. 11 maser spots used for an astrometric analysis are displayed with open circles and their absolute proper motion with arrows. Since they are close one another and they overlaps in this map.}\label{fig:f3}
\end{figure*}

The cross-power spectrum of the H$_2$O masers toward IRAS 06061+2151 from 2007/296 to 2009/250 with Mizusawa--Ishigakijima baseline (2270 km) are shown in Figure \ref{fig:f2}. The component around $v_{\mbox{\scriptsize LSR}} = -8$ km s$^{-1}$ showed the stable emission through the entire epoch. The systemic velocity ($v_{\mbox{\scriptsize sys}}$) of IRAS 06061+2151 is $-1.6\pm0.2$ km s$^{-1}$, which is obtained by \atom{CO}{}{12} observations of IRAS sources \citep{wouterloot89}. In total we detected 142 maser spots whose LSR velocities are ranged from $-$19.2 km s$^{-1}$ to 4.4 km s$^{-1}$, and their spatial distributions are shown in Figure \ref{fig:f3}. The maser spots are distributed over an area of 300 mas $\times$ 600 mas. Their distribution shows three groups, the northern, middle, and southern. The red-shifted components with respect to $v_{\mbox{\scriptsize sys}}$, are included in only the northern and middle groups, and blue-shifted components are concentrated in the southern group. This is consistent with the distribution reported by \citep{motogi08}. And these spots are located in \timeform{4".6} south-east of the peak position of sub-millimeter source (SCUBA 850 $\micron$ image was derived from {\it the Canadian Astronomy Data Centre}), and at the same position within \timeform{0".2} as the peak of the 8.4 GHz continuum source observed using VLA B-configuration \citep{kurtz94}.

\subsection{Measurements of the Annual Parallax}\label{parallax}
We used the 11 maser spots for a parallax measurement (see Sect. \ref{spotsize} for details). They were selected as they were detected during at least 6 epochs and their size was less than 2.5. The least square method was utilized for fitting the movement of each maser spot throughout the observing epochs to obtain the sinusoidal parallax curve and the linear proper motion represented by following equations \citep{green85};
\begin{eqnarray}
\alpha&=&\alpha_0+\mu_{\alpha}t+\pi F_{\alpha}\nonumber\\
\delta&=&\delta_0+\mu_{\delta}t+\pi F_{\delta}\nonumber
\end{eqnarray}
where, ($\alpha_0, \delta_0$) is the position at $t=0$, $\pi$ is the annual parallax, ($F_{\alpha}$, $F_{\delta}$) are sinusoidal functions of direction and time describing the parallax ellipse, and ($\mu_{\alpha}$, $\mu_{\delta}$) are proper motions in RA and Decl. respectively, $t$ is the interval in years from the standard epochs.

\begin{figure*}[t]
  \centering
    \FigureFile(140mm,){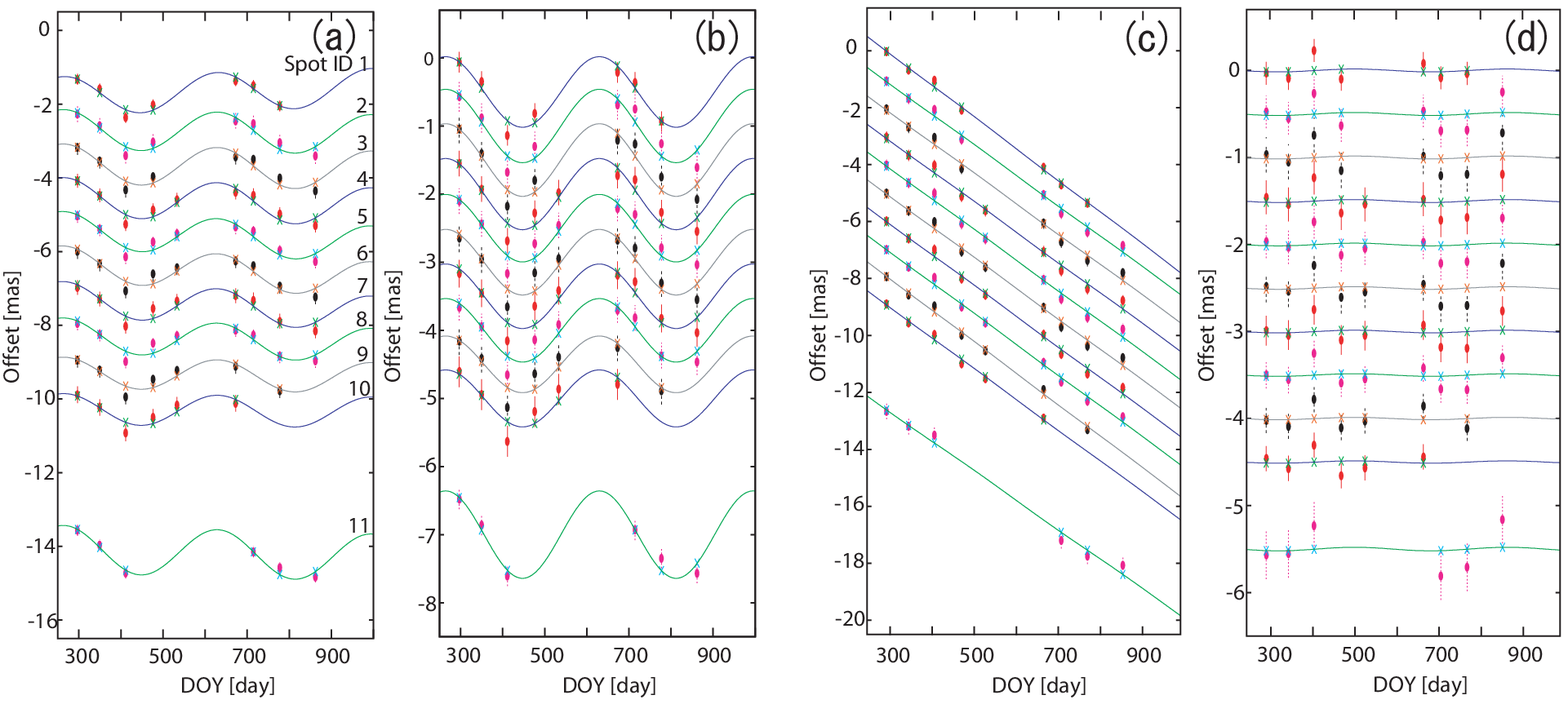}
  \caption{Movements of the 11 maser spots and the results of the least square fit for each maser spot in IRAS 06061+2151. Each panel shows as follows; (a) shows the displacement of RA offset against day of year (DOY) from 2007/250, (b) shows the displacement of RA offset removed the best fit proper motion from (a), (c) shows the displacement of Decl. offset against day of year (DOY) from 2007/250, and (d) shows the displacement of Decl. offset removed the best fit proper motion from (c), respectively. To clarify, the data for each spot were added small offset (-0.5 -- -32.0 mas) vertically. The Spot ID of each maser spots is labeled in Table \ref{tab:tbl2} and shown in the most left panel. The error bars represent the position uncertainties resulting from systematic errors which are given to make the reduced $\chi^2$ to be $\sim$ 1.}\label{fig:f4}
\end{figure*}

\begin{table*}[t]
\centering
\caption{The parallaxes and the absolute proper motions of each maser spot}
\label{tab:tbl2}

\scalebox{0.8}{
\begin{tabular}{ccccccccc} \hline
Spot ID & $v_{\mbox{\scriptsize LSR}}$ & \multicolumn{2}{c}{Position offset$^{\mbox{\scriptsize a}}$ [mas]} & Annual parallax$^{\mbox{\scriptsize b}}$ [mas] & \multicolumn{2}{c}{Absolute proper motion [mas yr$^{-1}$]} & Detected Epoch$^{\mbox{\scriptsize c}}$ & $N^{\mbox{\scriptsize d}}_{\mbox{\scriptsize epochs}}$ \\
 & & $\Delta\alpha\cos\delta$ & $\Delta\delta$ &  & $\mu_{\alpha}\cos\delta$ & $\mu_{\delta}$ &  &  \\\hline\hline
1 & -9.480 & -0.325(0.134) & -0.040(0.132) & $0.518\pm0.084$ & $0.11\pm0.12$ & $-4.04\pm0.10$ & 1111*111** & 7 \\
2 & -9.269 & -0.269(0.197) & -0.054(0.186) & $0.539\pm0.119$ & $-0.07\pm0.15$ & $-3.88\pm0.12$ & 1111*1111* & 8 \\
3 & -9.059 & -0.171(0.190) & -0.045(0.200) & $0.533\pm0.114$ & $-0.01\pm0.14$ & $-3.88\pm0.13$ & 1111*1111* & 8 \\
4 & -8.848 & -0.081(0.173) & -0.032(0.194) & $0.522\pm0.102$ & $-0.14\pm0.13$ & $-3.88\pm0.13$ & 111111111* & 9 \\
5 & -8.637 & -0.033(0.173) & -0.012(0.191) & $0.495\pm0.103$ & $-0.19\pm0.13$ & $-3.89\pm0.12$ & 111111111* & 9 \\
6 & -8.427 & -0.001(0.185) & -0.005(0.185) & $0.481\pm0.110$ & $-0.21\pm0.14$ & $-3.90\pm0.12$ & 111111111* & 9 \\
7 & -8.216 & 0.024(0.188) & 0.013(0.169) & $0.473\pm0.111$ & $-0.20\pm0.14$ & $-3.92\pm0.11$ & 111111111* & 9 \\
8 & -8.005 & 0.049(0.175) & 0.025(0.156) & $0.466\pm0.103$ & $-0.14\pm0.13$ & $-3.92\pm0.10$ & 111111111* & 9 \\
9 & -7.795 & 0.050(0.181) & 0.060(0.144) & $0.416\pm0.116$ & $-0.07\pm0.18$ & $-4.01\pm0.12$ & 111111*1** & 7 \\
10 & -7.584 & 0.114(0.196) & 0.119(0.145) & $0.419\pm0.137$ & $-0.05\pm0.27$ & $-3.92\pm0.18$ & 111111**** & 6 \\
11 & -7.373 & 18.435(0.129) & 12.334(0.276) & $0.641\pm0.120$ & $-0.12\pm0.12$ & $-3.75\pm0.19$ & 111***111* & 6 \\\hline
combined fit & & (0.181) & (0.183) & $0.496\pm0.031$ & & &  & -- \\\hline
\multicolumn{9}{@{}l@{}}{\hbox to 0pt{\parbox{210mm}{\footnotesize a: "Position offset" are offsets from the reference position which was set to be $\alpha=\timeform{06h09m06s.970593}$, $\delta=\timeform{+21D50'41".204897}$ (J2000.0). We determined this coordinate so that the position offset of the strongest maser spot in 2007/296 is (0, 0). Positions are those observed at the epoch detected for the first time, as noted in the "Detected Epochs". Numbers in parentheses represent the uncertainty.

b: The results were obtained from the RA--Decl. parallax fit.

c: Each digit shows the detected epochs of the maser spots corresponding to 2007/296, 2007/349, 2008/046, 2008/111, 2008/167, 2008/307, 2008/349, 2009/046, 2009/131, and 2009/250 respectively. "1" means that the maser spot was detected, and "*" means non-detection.

d: Total number of detected epochs.
}\hss}}
\end{tabular}}

\end{table*}

We conducted a combined parallax fit, in which the positions of 11 spots are fitted simultaneously with one common parallax but different proper motions and position offsets for each spot. As the result of this fit, the annual parallax of $0.496\pm0.031$ mas corresponding to a distance of $2.02^{+0.13}_{-0.12}$ kpc was obtained (see Table \ref{tab:tbl2}, Figure \ref{fig:f4}). Besides, the absolute proper motions of 11 maser spots are indicated in Figure \ref{fig:f3} as black arrows. In the fitting process, we obtained the position uncertainties (post-fit residuals) of 0.181 mas in RA direction, and 0.183 mas in Decl. direction, which are added to make the reduced $\chi^2$ to be approximately unity. The distance to IRAS 06061+2151 was determined directly with very high accuracy of 6.3 \%.

Here we prove that the error bars of 0.181 and 0.183 in RA and Decl., respectively are reasonable if we consider the uncertainties in astrometry. Possible errors in absolute position of the target are: a zenith atmospheric delay offset at each station, station position error (baseline error), and the thermal noise error. A zenith atmospheric delay offset at each station is caused by a difference in the optical path length ($\Delta l$) between two sources (target source--position reference source). The range of the elevation angle during our observations were between \timeform{30D} and \timeform{80D}. Given the mean zenith angle of observed sources of \timeform{55D}, the zenith atmospheric delay residual of 3 cm error, and separation of the zenith angles between two sources of \timeform{1.22D}, we can estimate $\Delta l$ to be 1.59 mm (Honma et al. 2007, 2008b; Nakagawa et al. 2008). The positional error of 144 $\mu$as due to this effect is derived from $\Delta l/B_{\mbox{\scriptsize max}}$, where $B_{\mbox{\scriptsize max}}$ is the maximum baseline length of VERA. The station position error was estimated to be 6 $\mu$as for a separation angle of \timeform{1.22D}, assuming that the position accuracy of 3 mm in each station, based on geodetic observations at the S and X bands \citep{jike09}. The thermal error due to noise was estimated to be 9 $\mu$as by dividing the half width of the synthesized beam size ($\theta_{\mbox{\scriptsize b}}$) by the SNR of phase referenced image. We used $\theta_{\mbox{\scriptsize b}}=1.8$ mas as the typical synthesized beam size (see Table \ref{tab:tbl1}) and SNR = 100 as the averaged SNR of phase referenced image of maser spots used for astrometric analysis. We note that the post-fit residual, which we obtained by the least-square fit of annual parallax is consistent with the sum of these estimated uncertainties.

\section{Discussion \& conclusions}

\subsection{Spot size}\label{spotsize}
In this section, we discuss the parameter "spot size" defined with VEDA (see in Sect. \ref{reduction}). The spot size of each maser spot depends not only on the maser structure but on the signal to noise ratio (SNR) of the brightest component in the map. Even if the maser spot has no structures, low SNR causes an increase of the spot size. In order to conduct an analysis of astrometric measurement with higher accuracy, the maser spots which have no structures and whose SNRs are substantially high, should be selected. By utilizing the maser spots which falls below the threshold for the spot size, it is possible to conduct an analysis of astrometric measurement with the maser spots which have no structures,  or are very bright.

In this paper, we used the maser spots which were detected at more than 6 epochs, and whose spot size is less than 2.5 for astrometric analysis. On the other hand, we also conducted a combined parallax fit for 13 maser spots which were also detected at more than 6 epochs, and their spot sizes were less than 5.0. As the result of the latter analysis, we obtained the annual parallax of $0.511\pm0.029$ mas. Comparing the former result with the latter, we found no significant difference. In the analysis of this study, there is little difference in the number of maser spots which were used for analysis with between former and latter conditions, because of the high threshold set for the number of detected epochs in which the maser spots were detected in this analysis. Although various values were also used to set the limit for the number of detected epochs and the spot size, we could not find the result that the structure of the maser spots adversely affects astrometric accuracy. The astrometric accuracy in present results of this source completely depends on the amount of statistics.

However, if the spot size of maser spot exceeds 3.0, several components appear around the brightest component as seen in the most right figure of Figure \ref{fig:f1}. In the case of this map, there would be no effect on the astrometric analysis because it is possible to distinguish the brightest component from other components (the brightest one exists as a compact component). But if some components associated with the brightest component are seen in the map, such a maser spot would adversely affect the astrometric result and its accuracy. So, we have to confirm the shape of maser spots in the phase referenced map to determine whether each maser spot has a structure which affects the astrometric result, or not. Only if we set a limit of 2.5 for the spot size such as in the present study, only compact maser spots are used for the astrometric analysis as seen in the left and middle maps of Figure \ref{fig:f1}.


\subsection{Motion of IRAS 06061+2151}\label{motion}

\begin{table*}[t]
\centering

\caption{The peculiar velocity of of IRAS 06061+2151}
\label{tab:tbl3}

\begin{tabular}{ccccccc}
\hline
\multicolumn{ 3}{c}{Galactic and solar motion parameters} &  & \multicolumn{ 3}{c}{Peculiar velocity} \\ \cline{1-3}\cline{5-7}
References of & $\Theta_0$ & $R_0$ &  & $U_p$ & $V_p$ & $W_p$ \\
solar motion parameters$^*$ & [km s$^{-1}$] & [kpc] &  & [km s$^{-1}$] & [km s$^{-1}$] & [km s$^{-1}$] \\ \hline\hline
\citet{dehnen98} & 220 & 8.5 &  & $9.6\pm0.7$ & $-28.5\pm2.3$ & $-11.6\pm1.6$ \\ 
 & $236\pm15$ & $8.0\pm0.5$ &  & $10.4\pm0.9$ & $-28.5\pm2.3$ & $-11.6\pm1.6$ \\ 
 &  &  &  &  &  &  \\ \hline
\citet{schonrich10} & 220 & 8.5 &  & $10.9\pm0.9$ & $-21.6\pm2.2$ & $-11.5\pm1.6$ \\
 & $236\pm15$ & $8.0\pm0.5$ &  & $11.7\pm1.1$ & $-21.6\pm2.2$ & $-11.5\pm1.6$ \\
 &  &  &  &  &  &  \\ \hline
Standard solar motion values & 220 & 8.5 &  & $10.2\pm0.6$ & $-18.5\pm2.2$ & $-11.0\pm1.5$ \\
 & $236\pm15$ & $8.0\pm0.5$ &  & $11.0\pm0.8$ & $-18.5\pm2.2$ & $-11.0\pm1.5$ \\
 &  &  &  &  &  &  \\ \hline
\multicolumn{7}{@{}l@{}}{\footnotesize *See Sect. \ref{motion}}
 
\end{tabular}
\end{table*}

The systematic motion of maser source is affected by its internal structure motions. In order to reduce their effects, we first calculated the un-weighted mean of the proper motions of the 11 maser spots in Table \ref{tab:tbl2}, and derived the mean proper motion to be ($\mu_{\alpha}\cos\delta$, $\mu_{\delta}$) = ($-0.10\pm0.10$, $-3.91\pm0.07$) mas yr$^{-1}$. Here, we calculated the standard deviation of the proper motions of 11 maser spots as the uncertainty of this mean proper motion. In order to obtain the proper motion in the Galactic coordinate, the tracking center position of IRAS 06061+2151 and the position added the mean proper motion values to the Galactic coordinate. We obtained the proper motion with respect to LSR in the Galactic coordinate to be ($\mu_l$, $\mu_b$) = ($3.37\pm0.01$, $-1.98\pm0.12$) mas yr$^{-1}$ by the difference of them. Given the source distance of 2.02 kpc, the proper motions ($\mu_l$, $\mu_b$) correspond to a velocity of ($v_l$, $v_b$) = ($32.20\pm0.12$, $-18.92\pm1.14$) km s$^{-1}$. Also, the systemic velocity of $-1.6\pm0.2$ km s$^{-1}$ derived from \atom{ CO}{}{12} observation was used as the radial velocity of this source.
 
In order to convert $v_{\mbox{\scriptsize LSR}}=-1.6\pm0.2$ km s$^{-1}$ to a heliocentric frame, $v_{\mbox{\scriptsize helio}}$, we used the following equation with the standard solar motion values ($U_{\solar}^{\mbox{\scriptsize std}}$, $V_{\solar}^{\mbox{\scriptsize std}}$, $W_{\solar}^{\mbox{\scriptsize std}}$) = (10.3, 15.3, 7.7) km s$^{-1}$ (see the appendix of Reid et al. 2009b).
\begin{eqnarray}
v_{\mbox{\scriptsize helio}} = v_{\mbox{\scriptsize LSR}}-(U_{\solar}^{\mbox{\scriptsize std}}\cos l+V_{\solar}^{\mbox{\scriptsize std}}\sin l)\cos b - W_{\solar}^{\mbox{\scriptsize std}}\sin b\nonumber
\end{eqnarray}
As the result of this calculation, we obtained $v_{\mbox{\scriptsize helio}}$ to be 10.8 km s$^{-1}$.

Next, to discuss about the Galactic 3-dimensional motions, we used the solar motion values of ($U_{\solar}$, $V_{\solar}$, $W_{\solar}$) = ($10.0\pm0.36$, $5.25\pm0.62$, $7.17\pm0.38$) km s$^{-1}$, derived from {\it Hipparcos} data by \citet{dehnen98}, $R_0$ = 8.5 kpc as the distance of the Sun to the Galactic center, and $\Theta_0 = 220$ km s$^{-1}$ as the rotation speed of the LSR. Also, we assume a flat rotation which means the rotation speed at the position of the target source $\Theta(R) = \Theta_0$, where $R$ is the distance from the target source to the Galactic center (in the case of IRAS 06061+2151, $R=10.5$ kpc). As a result of calculations using these observed and assumed parameters, we found that IRAS 06061+2151 implies peculiar motions of ($U_p$, $V_p$, $W_p$) = ($9.60\pm0.67$, $-28.50\pm2.26$, $-11.56\pm1.55$) km s$^{-1}$. where $U_p$ is peculiar motion of a star toward the Galactic center, $V_p$ is peculiar motion of a star in the direction of the Galactic rotation, and $W_p$ is peculiar motion toward the North Galactic Pole. With this calculation using the Galactic parameters assumed above, IRAS 06061+2151 showed a peculiar motion of larger than 28 km s$^{-1}$ counter to the Galactic rotation. In order to confirm whether this peculiar motion was caused by the model of the Galactic rotation or not, we re-calculated the peculiar motion of IRAS 06061+2151 by using the solar motion of ($U_{\solar}$, $V_{\solar}$, $W_{\solar}$) = (10.3, 15.3, 7.7) and ($11.10^{+0.69}_{-0.75}$, $12.24^{+0.47}_{-0.47}$, $7.25^{+0.37}_{-0.36}$) km s$^{-1}$ \citep{schonrich10}, $R_0 = 8.0\pm0.5$ kpc \citep{reid93}, and $\Theta_0 = 236\pm15$ km s$^{-1}$ \citep{reid04}. We list the result of re-calculation of the peculiar velocity components in Table \ref{tab:tbl3}. Though the Galactic parameters were variously changed, the motion of IRAS 06061+2151 counter to the Galactic rotation was larger than 17.0 km s$^{-1}$. Therefore, we suggest that IRAS 06061+2151 orbits the Galaxy slower than its rotation.

The peculiar motions of several sources which are located in the Perseus spiral arm are shown in Table \ref{tab:tbl4}. The parallaxes and the proper motions of these sources have already been measured by previous studies. We calculated ($U_p$, $V_p$, $W_p$) of each source using $R_0 = 8.5$ kpc, $\Theta_0 = 220$ km s$^{-1}$ , ($U_{\solar}, V_{\solar}, W_{\solar}$) = ($10.0\pm0.36$, $5.25\pm0.62$, $7.17\pm0.38$) km s$^{-1}$ and the parameters of each source already reported \citep{moellenbrock09, moscadelli09, oh10, reid09a, sato08, xu06}. We used the systemic velocities derived from CO and HCO$^{+}$ observations \citep{torrelles87, wouterloot89, molinari02, lee03} as the radial velocity for H$_2$O maser sources, because the line width of H$_2$O maser sources is generally very broad (from a few tens of km s$^{-1}$ to 100 km s$^{-1}$). Six sources (IRAS 06061+2151, IRAS 06058+2138, AFGL2789, W3OH, NGC7538, IRAS00420+5530) orbit the Galaxy at least 15 km s$^{-1}$ slower than the expected Galactic rotation, and they move at on average 27 km s$^{-1}$ toward the Galactic center and counter to the Galactic rotation.

\begin{table*}[t]
\centering

\caption{The peculiar motions of seven sources in the Perseus spiral arm}
\label{tab:tbl4}

\scalebox{0.9}{
\begin{tabular}{lcrrrrrrc} \hline
\multicolumn{1}{c}{Source} & $D$ & \multicolumn{1}{c}{$l$} & \multicolumn{1}{c}{$b$} & \multicolumn{1}{c}{$U_p$} & \multicolumn{1}{c}{$V_p$} & \multicolumn{1}{c}{$W_p$} & maser emission & Ref.$^{\mbox{*}}$ \\
\multicolumn{1}{c}{} & [kpc] & \multicolumn{1}{c}{[deg]} & \multicolumn{1}{c}{[deg]} & [km s$^{-1}$] & [km s$^{-1}$] & [km s$^{-1}$] &  &  \\\hline\hline
IRAS 06061+2151 & $2.05^{+0.13}_{-0.12}$ & 188.79 & 1.03 & $9.6\pm0.7$ & $-28.5\pm2.3$ & $-11.6\pm1.6$ & H$_2$O & This paper \\
IRAS 06058+2138 & $1.76^{+0.11}_{-0.11}$ & 188.95 & 0.89 & $3.9\pm1.0$ & $-21.3\pm6.2$ & $4.0\pm4.4$ & H$_2$O & 1 \\
 & $2.10^{+0.03}_{-0.03}$ &  &  & $-4.4\pm3.0$ & $-15.7\pm0.9$ & $-2.0\pm0.5$ & CH$_3$OH & 2 \\
AFGL2789 & $3.07^{+0.30}_{-0.30}$ & 94.60 & -1.80 & $14.0\pm3.3$ & $-33.5\pm3.3$ & $-11.0\pm4.4$ & H$_2$O & 1 \\
W3OH & $1.95^{+0.04}_{-0.04}$ & 133.95 & 1.06 & $22.8\pm2.4$ & $-17.9\pm2.0$ & $1.0\pm0.4$ & CH$_3$OH & 3 \\
NGC7538 & $2.65^{+0.12}_{-0.11}$ & 111.54 & 0.78 & $24.2\pm1.9$ & $-29.4\pm2.8$ & $-10.9\pm1.1$ & CH$_3$OH & 4 \\
IRAS 00420+5530 & $2.17^{+0.05}_{-0.05}$ & 122.02 & -7.07 & $25.5\pm2.1$ & $-20.6\pm2.3$ & $6.1\pm0.7$ & H$_2$O & 5 \\
NGC281 & $2.82^{+0.24}_{-0.24}$ & 123.07 & -6.31 & $11.4\pm3.9$ & $-0.7\pm4.2$ & $-13.6\pm2.5$ & H$_2$O & 6 \\\hline
\multicolumn{9}{@{}l@{}}{\footnotesize *References are 1:\citet{oh10}, 2:\citet{reid09a}, 3:\citet{xu06}, 4:\citet{moscadelli09}, 5:\citet{moellenbrock09}, 6:\citet{sato08}}

\end{tabular}}

\end{table*}

We note that NGC 281 does not show this peculiar motion, this would be because that NGC 281 is slightly offset from the Galactic plane and is associated with an expanding super-bubble \citep{sato08}. Thus, the result of our observation of IRAS 06061+2151 would be consistent with the conclusion that previous studies imply a non-circular rotation of the Perseus spiral arm.


\bigskip

We thank an anonymous referee for careful reading, and valuable comments. We are also extremely grateful to the staff of all the VERA stations for their assistance in observations. This research used the facilities of the Canadian Astronomy Data Centre operated by the the National Research Council of Canada with the support of the Canadian Space Agency, and the SIMBAD database operated at CDS, Strasbourg, France.




\end{document}